\documentclass[aps,pra,twocolumn,showpacs,groupedaddress]{revtex4}
\usepackage{graphicx}

\begin{document}

\title{Thermal emissivity for finite three-dimensional photonic band gap crystals}

\author{Andrew J. Stimpson}
\email[Electronic Address: ]{Andrew.Stimpson@jpl.nasa.gov}
\author{Jonathan P. Dowling}
\affiliation{Quantum Computing Technologies Group\\
Exploration Systems Autonomy, Section 367\\
Jet Propulsion Laboratory, California Institute of Technology\\
Mail Stop 126-347, 4800 Oak Grove Drive, Pasadena, California
91109}

\date{\today}

\begin{abstract}
We discuss the results of computer model for the thermal emissivity of a three-dimensional photonic band gap (PBG) crystal, specifically an inverted opal structure.  The thermal emittance for a range of frequencies and angles is calculated.
\end{abstract}


\maketitle

\section{INTRODUCTION}
It has been shown that there exist periodic dielectric structures such that
for a range of frequencies, no electromagnetic waves can propagate in any
direction for all polarizations~\cite{old1,old2,old3,old4,old5}. The
properties of such an infinite crystal have been worked out in
detail~\cite{john-wang90, john-wang91}. Such work is typically limited to
non-absorptive materials, since otherwise all waves become evanescent.

Much theoretical work has been done on the properties of finite one
dimensional photonic band gap (PBG) crystals~\cite{bendickson96}, including recent calculations of the thermal emissivity of such one-dimensional structures~\cite{cornelius99,dowling01}.  The strong angular dependence of the gap effect with a one-dimensional structure has motivated successful experimental work with three-dimensional structures~\cite{shawn-yu00}.  The three-dimensional model we will use here is an inverted opal structure, and with the use of a computer program published by Stefanou et al.~\cite{multem1,multem2}, the transmittance, reflectance, and absorbance of a finite number of layers of the crystal is calculated, as well as the complex band structure of the corresponding infinite crystal.  As we shall show, our structure exhibits a broad band gap in the emissivity that has virtually no angular dependence, and as such, also suppresses the thermal emittance of the crystal in an omnidirectional fashion.  We also predict emissivity enhancement at the photonic band edge.

For a given crystal structure, we can calculate ${\cal T}$ and ${\cal R}$,
the transmission and reflection coefficients of the crystal as a function of the frequency $\omega$ and the wave vector ${\mathbf k}$ of the incident wave. If we consider a crystal with real indices of refraction, then ${\cal R} + {\cal T} = 1$ from conservation of energy.  However, if there is an imaginary part to the any of the indices of refraction, there will be a non-zero absorbtion coefficient ${\cal A} = 1 - {\cal R} - {\cal T}$. Calculation of this quantity is useful because of a Kirchoff's second law which states that a material's thermal emmittance ${\cal E}$ is proportional to its absorbance (and that they are equal for a blackbody).  One can obtain the absorbance by simply multiplying Planck's blackbody power spectrum by the absorbtion coefficient.  So by modelling the absorbtion of several layers of a three-dimensional crystal coated on the surface of a thick, highly absorptive substrate, we are also modelling the thermal emissivity $E$ of the entire structure.  The three-dimensional superlayer acts as a passive filter, which alters the emissivity of the underlying graybody substrate.

\section{THEORY}
\subsection{One-dimensional case}
We employ software developed by Stefanou et al.~\cite{multem1,multem2} that uses a generalization of the matrix-transfer techniques developed for one-dimensional crystals, such as thin-film dielectric stacks. In the one-dimensional case, the ${\cal T}$ and ${\cal R}$ are $ \left| t \right|^2$ and $ \left| r \right|^2$ respectively, where $\textsl{t}$ and $\textsl{r}$ are the complex transmittance and reflectance coefficients, respectively.  These coefficients are linearly related to one another by a $2 \times 2$ matrix $\mathbf{\hat{M}}$, where
\begin{equation}
\left[ \begin{array}{c}
1 \\
r \end{array} \right] = \mathbf{\hat{M}} \left[ \begin{array}{c}
t \\
0 \end{array} \right] \label{1d_linear:1} \end{equation}
and
\begin{equation}
\mathbf{\hat{M}} = \left[ \begin{array}{cc} M_{11} & M_{12}\\
M_{21} & M_{22} \end{array} \right]. \label{1d_linear:2}
\end{equation}

The column vectors on each side of Eq. (\ref{1d_linear:1}) represent the
fields outside each side of the crystal.  Then $t = 1/M_{11}$ and $r =
M_{21}/M_{11}$.  This matrix $\mathbf{\hat{M}}$ can be constructed for a one-dimensional structure simply by multiplying the transfer matrices for the individual layers.  For a homogenous plate, these can be readily computed by taking into account the the reflection and transmission at the interface at the edges of the plate with the Fresnel equations, along with the accumulated phase (and possible change in amplitude if the index of refraction has an imaginary component) from propagating though the material.

Because the phase accumulated from the wave propagating though a given layer
depends on the portion of the wave-vector that is perpendicular to the
interfaces of the plates, the frequency $\omega$ at which the gap is centered is highly dependent on the angle of incidence of the initial radiation.  As a consequence, while the on-axis thermal emissivity at a given $\omega$ might be suppressed, off-axis emissivity might have a band-edge enhancement, as shown in earlier work~\cite{cornelius99,dowling01}.

\subsection{Three-dimensional model}
The model used by the Stefanou and co-workers~\cite{multem1,multem2} also allows for a finite number of parallel planes of spheres of differing index of refraction from the host material in which they are embedded.  For a given frequency $\omega$, the incident plane wave is expanded into
spherical waves about an origin centered at one sphere, and then the
scattered wave from a single sphere in one plane is calculated by satisfying
the boundary conditions for Maxwell's equations at the surface of a single
sphere.  The scattered wave for the entire infinite plane of spheres is just
the sum of the scattered waves for a single sphere expanded about a
translated origin, with a phase factor corresponding to the portion of the
incident wave vector that is parallel to the plane of spheres.  By summing
this first-order scattering of all the spheres, \emph{except} the sphere at
the origin, and then expanding the result into spherical waves about the
origin, we can get a better approximation of the true wave that is incident
on the sphere at the origin.

Here, the restriction that each plane of spheres has the same two-dimensional periodicity allows for the difficult part of this expansion to be calculated only once per frequency, which is what makes this method much more numerically feasible than other approaches.

By solving the boundary conditions again, with the spherical wave expansion of the incident plane wave plus this first order scattered wave, we solve for a second-order scattering off the sphere at the origin.  By symmetry, all the spheres of the infinite plane have this same scattered wave.  Hence, the total scattered wave for the plane is again the sum of the scattered waves for a single sphere, expanded about a translated origin, with
a phase factor.  The end result is then expanded into plane waves and added
to the incident wave to to solve for the total transmitted and reflected field.  By repeating this process for each plane of spheres, the total field for the entire crystal is obtained.

\section{RESULTS}

\begin{figure}
\includegraphics[width=7cm]{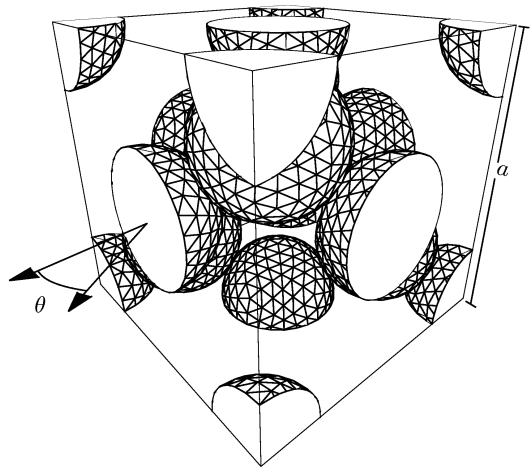}
\caption{\label{unitcell}A unit cell of the inverted opal structure used.}
\end{figure}

\begin{figure}

\includegraphics[width=7cm]{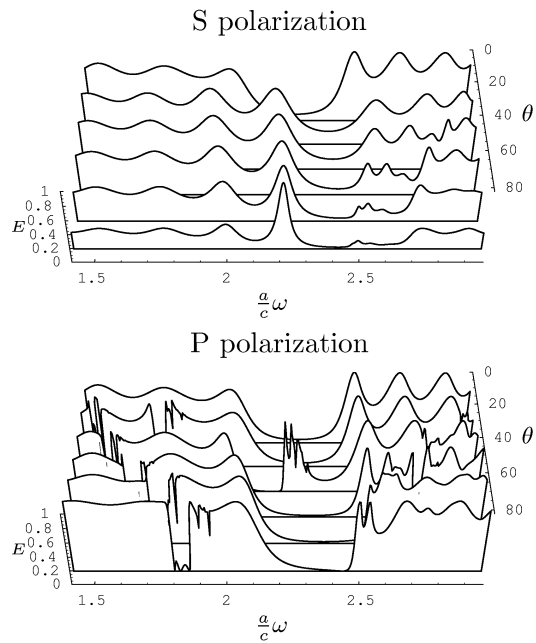}

\caption{\label{3d}The angular dependence of the emissivity for the inverted opal structure with four periods.  The frequency is given in units of $c / a$, where $c$ is the vacuum speed of light and $a$ is the length of the unit cell in Fig.~\ref{unitcell}.}

\end{figure}

\begin{figure}

\includegraphics[width=7cm]{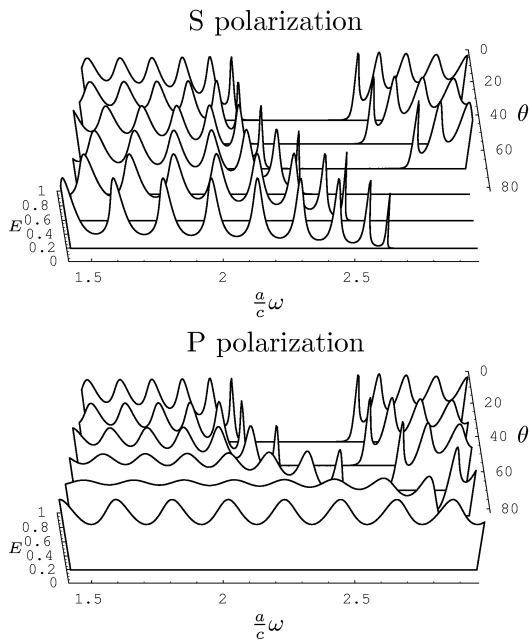}

\caption{\label{1d}Emissivity for a one-dimensional crystal with 16 periods at several angles from normal.  Contrasted with a three-dimensional crystal, it has a much stronger angular dependence.}
\end{figure}

\begin{figure}
\includegraphics[width=7cm]{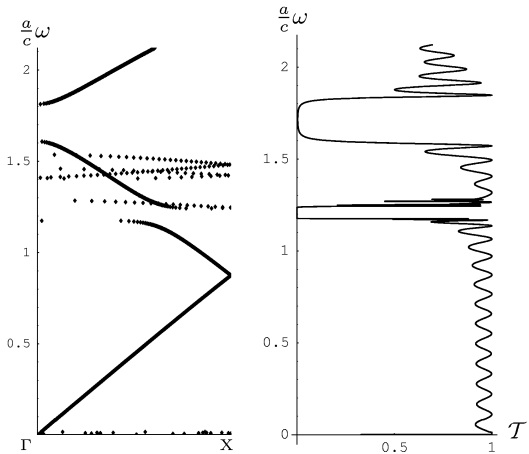}
\caption{\label{band}Part of the band structure (left) and transmission (right) of the infinite, non-absorbing crystal with $\epsilon_1 = 22$.}
\end{figure}

The three-dimensional crystal chosen for our calculations is a diamond fcc inverted opal (Fig.~\ref{unitcell}) with the complex relative dielectric constant $\epsilon_1 = 12 + i/10$ for the space around the spheres and $\epsilon_2 = 1$ inside the spheres, which are of radius $0.30618621 \, a$.  Four layers of the unit cell are stacked on a highly absorptive substrate ($\epsilon = 12 + 7i$) which is eight orders of magnitude thicker than the height of the total crystal, so that it can be considered essentially infintely thick.  The emissivity dependence on normalized frequency and polar angle for both polarizations is given in Fig.~\ref{3d}.

Several important features are noted.  Firstly, there is a large band-gap in the emissivity that is largely angle independent.  Secondly, the emissivity of the graybody substrate goes exactly to unity at the band-edge resonance frequencies.  This means that the emission is enhanced up to the pure blackbody limit rate at these frequencies.  This observation is consistent with our previous one-dimensional calculations~\cite{cornelius99,dowling01}, and recent experimental observations~\cite{shawn-yu00}.  As noted in our previous work, around the band-edge resonances the three-dimensional photonic crystal superstrate acts as an antireflective coating, allowing all incident radiations at the frequency to be absorbed---which is the definition of a blackbody.

You will notice that a complete gap does not exist for this structure, as is the case for most inverted opals, as you can see emission is not suppressed for a particular angle for the center gap frequency of $2.27 \, c/a$.  While other structures might show more promise of having complete stop bands, opals are unique in that large crystals can be grown by self-assembly of individual spherical particles, which can then be doped out to provide the inverted opal structure we have presented.

A complementary set of plots for a one-dimensional crystal is given in Fig.~\ref{1d}.  For the one-dimensional crystal, indices of refraction and thicknesses chosen such that the absorption gap is centered at the same frequency and of the same width as the three-dimensional crystal.  This is achieved by setting $\epsilon_1 = 2.6$, $d_1 = 0.6 \, a$, $\epsilon_2 = 1.44$, and $d_2 = 0.81 \, a$ where $d_1$ and $d_2$ are the thicknesses of the the two layers and $\epsilon_1$ and $\epsilon_2$ are the corresponding relative dielectric constants.  Here, $a$ is once again the length of the edge of the unit cell of the three-dimensional crystal, so all frequencies are comparable.

While the frequencies corresponding to suppressed emission in the one-dimensional crystal  are highly dependent on angle, the emissive gap effect in the three-dimensional crystal is nearly independent of angle.  Though, as mentioned before, a band structure cannot be calculated for a finite, absorptive structure, one can calculate the band structure for the infinitely extended crystal with the absorption removed by setting $\mbox{Im}(\epsilon_1) = 0$ and removing the backplane.  However, the existence of the gap effect in our configuration is highly dependent on the boundary condition of the thick absorptive backplane and the presence of absorbtion in the crystal itself, and no gap effect is present if they are removed.  This geometry does exhibit gap behavior at a different frequency, though, if $\epsilon_1$ is changed appropriately (Fig.~\ref{band}).

\section{CONCLUSION}

We have calculated the thermal emissivity of a three-dimensional PBG crystal using extensions of tested numerical techniques developed for 1D crystals.  Our matrix-transfer approach is conceptually much simpler than the traditional QED method of solving the non-trivial boundary conditions involved, and also not nearly as computationally intensive as the Finite Difference Time Domain (FDTD) simulations used by many groups for solving Maxwell's equations in PBG crystals.  Our technique also takes into account the essentially finite nature of the structure, crucial for experimental work where often it is possible to fabricate only a few layers of crystal.  Our results agree qualitatively with recent experiments by S. Lin~\cite{shawn-yu00}.

\section*{ACKNOWLEDGEMENTS}

This work was performed at the Jet Propulsion Laboratory, California Institute of Technology, under a grant from the National Aeronautics and Space Administration.  Additional support was provided by the National Reconnaissance Office, the National Security Agency, the Advanced Research and Development Activity, the Office of Naval Research, and the Defense Advanced Research Projects Agency.

\bibliography{draft3}

\end{document}